\documentclass[]{spie}  

\usepackage{amsmath,amsfonts,amssymb}
\usepackage{graphicx}
\usepackage[colorlinks=true, allcolors=blue]{hyperref}
\usepackage{booktabs}
\usepackage{xcolor} 

\usepackage[utf8]{inputenc}
\usepackage{graphicx}
\usepackage{subcaption}
\title{Rubin M1M3 Dynamic performance – stability and actuation during operations}

\author[a]{HyeYun Park}
\author[b]{Petr Kubánek}
\author[b]{Kshitija Kelkar}
\author[c]{Ignacio Sevilla-Noarbe}

\author[f]{Andrea Jeremie}
\author[d]{Brian Stalder}
\author[b]{Bruno C. Quint}
\author[b]{David Sanmartim}
\author[f]{Dominique Boutigny}
\author[d]{Douglas R. Neill}
\author[d]{Erik Dennihy}
\author[b]{Felipe Daruich}
\author[b]{Freddy Mu\~noz Arancibia}
\author[e]{Kevin Fanning}
\author[e]{Kevin Reil}
\author[c]{Laura Toribio San Cipriano}
\author[d]{Malhar Sonaniskar}
\author[b]{Marina S. Pavlovic}
\author[g]{Noah Gonzalez}
\author[b]{Paulina Venegas}
\author[d]{Sandrine Thomas}
\author[d]{Tiago Ribeiro}
\author[e]{Yijung Kang}

\affil[a]{Duke University, 120 Science Drive, Durham, USA}
\affil[b]{NSF-DOE NOIRLab, 1500 Av. Juan Cisternas, La Serena, Chile}
\affil[c]{Centro de Investigaciones Energ\'eticas, Medioambientales y Tecnol\'ogicas, CIEMAT, Av. Complutense 40, Madrid, Spain}
\affil[d]{NSF-DOE Vera C. Rubin Observatory Project Office, Tucson, AZ, USA}
\affil[e]{SLAC National Accelerator Laboratory, 2575 Sand Hill Rd., Menlo Park, CA 94025, USA}
\affil[f]{LAPP, Université Savoie Mont Blanc, CNRS/IN2P3, Annecy; France.}
\affil[g]{Instituto de Física y Astronomía, Universidad de Valparaíso, Av. Gran Bretaña 1111, Playa Ancha, Chile}

\authorinfo{Further author information: (Send correspondence to HyeYun P.)\\HyeYun P.: E-mail: hyeyun.park@duke.edu, Telephone: 1 631 384 5240}

\begin{document} 
\maketitle

\begin{abstract}
The Vera C. Rubin Observatory is preparing to commence the Legacy Survey of Space and Time with the fully integrated Simonyi Survey Telescope. To verify that the primary/tertiary (M1M3) mirror system is ready to meet the demanding survey requirements, dynamic tests of the 8.4 m, 53 ton M1M3 system were conducted to assess safety, stability, and image quality under realistic operating conditions. The M1M3 is supported by 156 pneumatic force actuators and positioned, relative to its mirror cell, by six hardpoint actuators that together must counteract gravitational and inertial loads during rapid telescope motion. The Rubin Observatory telescope mount is capable of moving at a rate that meets its nominal motion requirements (Az: vel +/- 7 deg/sec, accel +/- 7 deg/sec2 / El: vel +/- 3.5 deg/sec, accel +/- 3.5 deg/sec 2), and can approach it maximum allowable values that are 50 percent higher. Even at just 20\% of its operational speed, it is an exceptionally fast motion for such a large structure. After slewing, the system must stabilize and dampen vibrations within 5 seconds to ensure image quality during observations. Achieving this rapid settling requires precise control of 156 force actuators, which must adjust dynamically with changes in telescope elevation to compensate for gravity effects. We present results for M1M3 from a comprehensive series of TMA dynamical tests spanning the operational envelope of slew velocities and accelerations. The analysis evaluates elevation axis balancing and lookup table updating as we install the M1M3 mirror; slew-and-settle behavior, force response and stability of the pneumatic actuator system across telescope attitudes including responses to the earthquake. The results demonstrate the readiness of the M1M3 subsystem for routine survey operations and provide validation data for ongoing performance modeling.
\end{abstract}


\section{INTRODUCTION}
\label{sec:intro}  
The Vera C. Rubin observatory\cite{Ivezic} has been going through milestones to get ready to start the 10 year survey. From October to December 2024, we had the On-sky commissioning with the Commissioning Camera (ComCam)\cite{comcam}, which ended with successful verification of system functionality.  In March 2025, the LSST Science Camera\cite{LSSTCam} was installed on the telescope. That was followed by the LSSTCam commissioning started April 2025, which includes the first ``test shots" with the full camera. On June 23, 2025 ``First Look" images were officially released to the public. October 25, 2025 Official handover from the Construction team to the Operations team, and it's on the final stage to start the 10-year LSST Survey.

The previous report, Rubin M1M3 Support System Dynamic Performance \cite{10.1117/12.3019268}, showed that the M1M3 inertial compensation system is capable to compensate inertial forces at up to 70\% of the telescope's maximum velocity, acceleration and jerk. However, that report was built on data collected using a mirror mass surrogate. The M1M3 glass mirror was installed on the Telescope Mount Assembly (TMA) later in October 2024. It enabled subsequent testing on sky with the Commissioning Camera (ComCam), which was already installed on the telescope for our first on-sky tests from October until December, 2024. Later, in March 2025, the on-sky commissioning tests continued after the installation of the LSSTCam. This study includes all the dynamic performance tests and their results since October 2024 to April 2026. 

\begin{figure}
    \centering
    \includegraphics[width=0.8\textwidth]{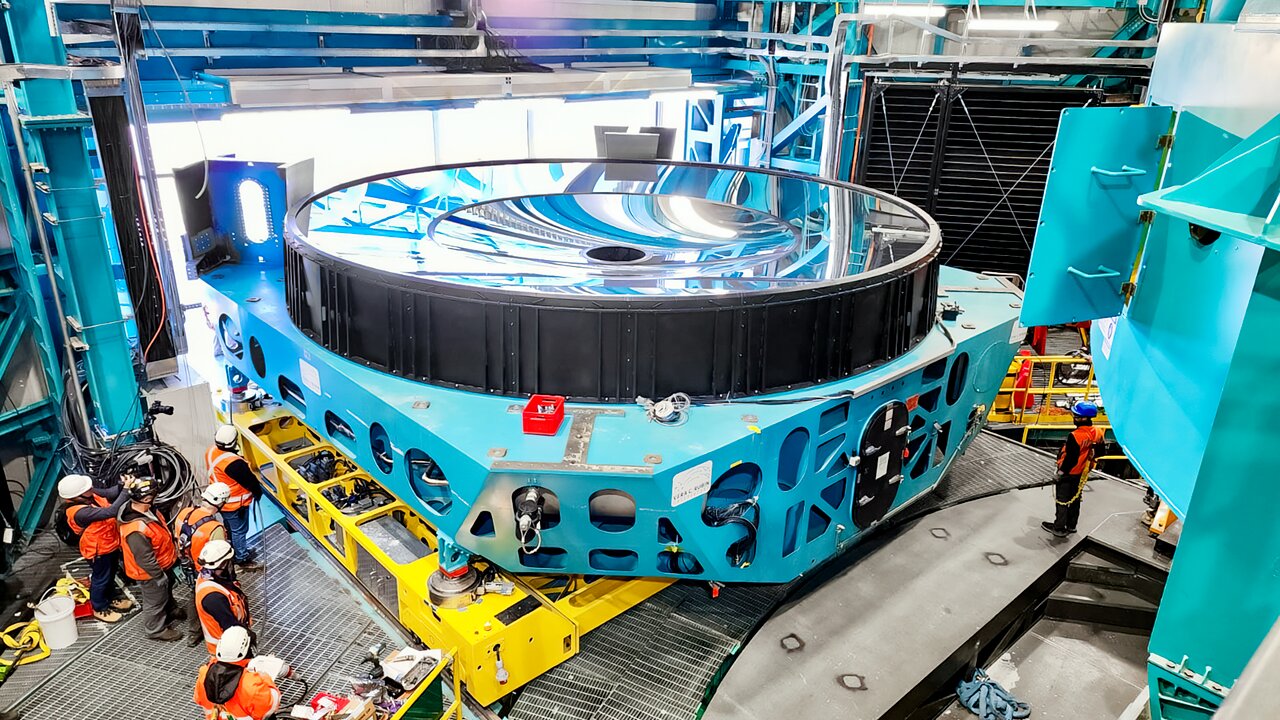}
    \caption{M1M3 insgtallation. RubinObs/NOIRLab/SLAC/NSF/DOE/AU.}
    \label{fig:m1m3_install}
\end{figure}


M1M3\cite{M1M3final} consists of 156 pneumatic force actuators (FA)\cite{FA} and six hardpoints (HP)\cite{hardpoints_2018, hardpoints_2008}. 44 of the force actuators are single-axis and 112 of them are dual-axis. Single axis FAs support the mirror in the axial direction only, and dual-axis FAs support lateral direction or cross-lateral direction in addition to the axial direction. 

Not only are the force actuators supporting the mirror, they are also counteracting inertia forces and adjusting the optical figure for the Active Optics System (AOS) for better image quality. The gravity force at each elevation is compensated with actuator forces provided by a look-up table (LUT). Section \ref{sec:lut}) contains details about LUT updates. Then for safety, dynamic tests are done at very low speeds of the TMA, and the speed is slowly  increase step by step. The M1M3 mirror has a radius of 8.4 m and a weight of ~17 tons. The total weight of the M1M3 mirror, its support system, its thermal control system and its mirror cell is over 50 tons. One of the important requirement of the massive TMA (Section \ref{sec:slew_settle}), is that it needs to slew and settle in a very short time limit. The TMA should also not over shoot after the emergency brake (Section \ref{sec:braking_distance}), and mirrors should return back to the same positions, relative to their mirror cells, after the slew.  The observers run the quasi-static tests for the FAs and HPs daily, before raising the mirror, to check their functionality for the safety of the mirror (Section \ref{sec:quasi-static}). 
Since the Rubin observatory is located in a high seismicity region, the building itself, interlock system, and mirror systems are designed to safely react to earthquake events (Section \ref{sec:earthquake}). 

\section{BALANCING AND LOOKUP TABLE}\label{sec:lut}
M1M3 mirror cell assembly has pneumatic Force Actuators (FAs) for supporting the M1M3 mirror and hardpoint actuators (HP) for controlling its position, relative to its mirror cell. The FAs must counteract gravitational and inertial loads during rapid telescope motion. The HPs only measure the forces that are transmitted through them. These forces are counteracted by the FAs, through a force balance system.

To ensure the structural integrity and optical precision of the Vera C. Rubin Observatory’s massive monolithic mirror, the LSST M1M3 support system\cite{M1M3CSC} must effectively decouple global gravitational loads into specific demands for each of its 156 FAs. When the telescope tilts, the total weight of the mirror redistributes between the axial and lateral supports based on the elevation angle.

The gravity LUT (G-LUT) values were initially determined from the M1M3 Finite Element Model\cite{LSSTModel}. After the glass M1M3 was installed on the TMA, the G-LUT was refined using excess forces captured by the M1M3 force balancing system. This was only possible when the M1M3 glass was tilted in elevation. The first time the M1M3 was tilted, it was only tilted 10 degrees. This was for in factory tests using a steel surrogate mirror. This setup was principally for force actuator testing. Only the Telescope Mount Assembly (TMA)\cite{TMA} enables unrestricted motion of the M1M3 glass mirror.

The Large Binocular Telescope (LBT)\cite{LBTM1}, the only other operational telescope using the 8.4m University of Arizona mirrors, relies on sine and cosine decoupling distributing forces to all its force actuators. Simonyi's telescope support system design uses a bit different approach - the system employs a 5th-degree polynomial expansion calculated from the TMA elevation. This polynomial serves as a fine enough approximation of the sine and cosine functions naturally required to distribute forces on a tilted plane.

Using polynomials instead of geometric functions enables higher flexibility. The extra flexibility is needed as the M1M3 is different from LBT's M1 - it combines two optical surfaces with a different thickness into one big mirror. This approach also offers possibility to fine-tune FA specific forces as needed - something that isn't possible with a simpler geometric functions-based decoupling. The system accurately compensates for gravity’s vector, ensuring that each FA applies the specific forces needed to maintain the mirror’s figure and position regardless of where the telescope is pointing in the sky.

\begin{figure}[htbp]
    \centering
    \includegraphics[width=1\linewidth]{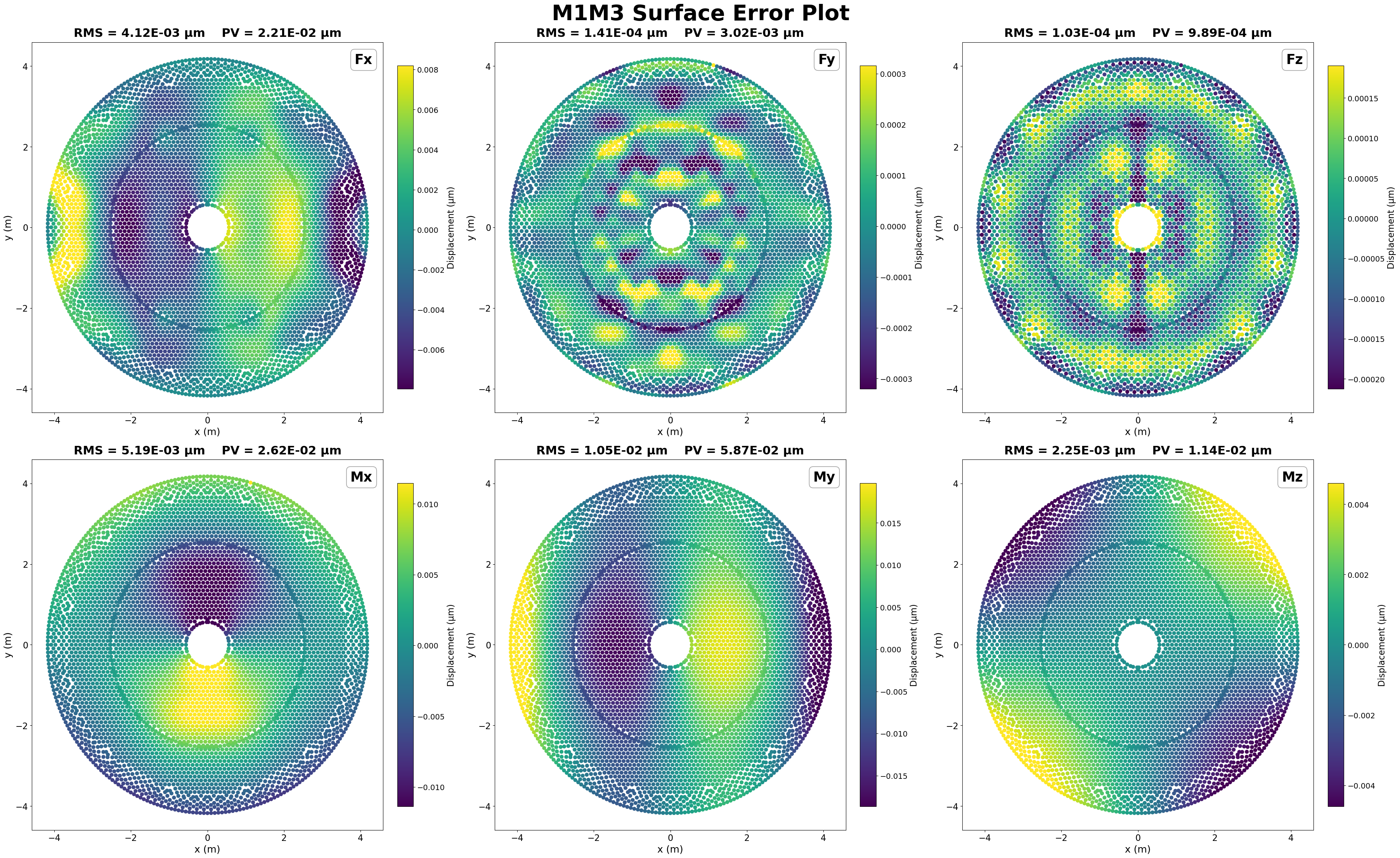}\caption{M1M3 balance forces - surface error plots}\label{fig:m1m3balancing}
\end{figure}

The dynamic correction forces are composed of six independent balance force sets, corresponding to corrections in Fx, Fy, Fz, Mx, My, and Mz at the mirror Center of Gravity (C.G.) as shown in Figure~\ref{fig:m1m3balancing}. Each force set should produce net zero reaction forces at the six hardpoints. These balance force sets were determined using finite element analysis (FEA), which was also used to quantify the optical surface error and mirror stresses. The load is applied to the FEA model, and the resulting optical surface error determined, along with the resulting actuator forces. The optical surface is then decomposed into bending modes. The actuator forces corresponding to the bending modes are then combined with the previously FEA determined actuator forces to develop an actuator force set that minimizes the surface error\cite{m1m3_bending}. This LUT matrix has been refined several times based on measured optical errors. The actual constraint reactions at the hardpoints were at least three orders of magnitude smaller than the applied forces. Additionally, during slewing any unintended loading of hardpoints would affect force measurements used by the safety interlock system and potentially trigger a fault. During observing, load cells in each hardpoint measure the load transmitted through them. Similarly to the dynamic correction the control system then applies loads distributed to the FA to minimize these hardpoint loads. 

This optimization reduced the Fx correction surface error from 206 nm to 22.1 nm PV (34.3 nm to 4.12 nm RMS) and the Mx correction from 234 nm to 26.2 nm PV (47.6 nm to 5.2 nm RMS). The maximum principal stress within mirror glass, peaked at 15 kPa (for Fx) across all sets, confirming mirror safety limits. The resulting force sets form the basis of updated LUTs. 

Telescope balancing has been done every time a new payload is installed on the TMA. The mandatory prerequisite for balancing is that telescope is well within the coarse balance range\cite{balancing}, below 500 Nm or the residual torque on elevation axis when moving full range from 0 to 90 deg in elevation\cite{10.1117/12.3019266}. After the telescope balancing, the LUT was updated based on the measured actuator and hard point forces, as a function of elevation angle. 

\section{DYNAMIC TESTS WITH DIFFERENT TMA SPEED SETTINGS}

Thanks to the campaign of tests performed with the surrogate cell \cite{10.1117/12.3019268} we had confidence that we could proceed to pre-operations with TMA settings that would mimic the regular `Feature Based Scheduler' (FBS)-driven slews  during operations (the current pre-LSST early on-sky operations with $20\%$ of the maximum design speed, acceleration and jerk of the mount assembly). We report here on the results of mirror positioning tests for a full night of observations, covering the slew and settle time of M1M3.


\subsection{Slew and Settle time}\label{sec:slew_settle}

For a short slew (3.5 degrees on the sky in any direction), the entire TMA and optical systems is required to slew and settle to the same position (within precision) within 5 seconds. The TMA itself is required to slew and settle in 4 seconds, allowing 1 second for the optical systems to converge.


The location of the M1M3 relative to its mirror cell is determined by an Independent Measurement System (IMS)\cite{ims} composed of a set of electronic micrometers that measure the displacement of the M1M3 mirror with respect to the cell. 

\begin{figure}[htbp]
    \centering
    \includegraphics[width=0.65\linewidth]{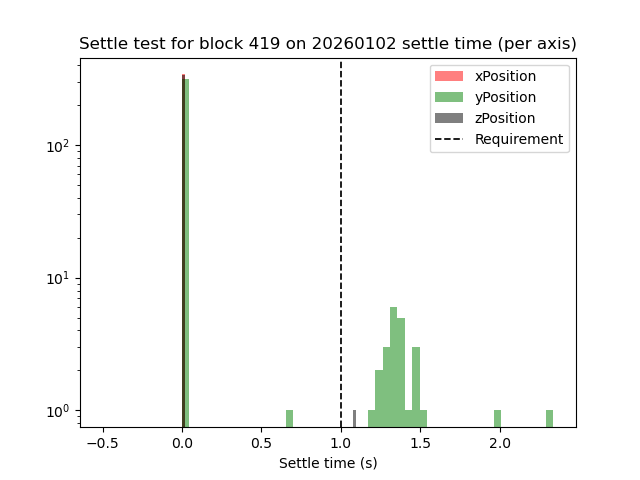}\caption{M1M3 settle time in axes $xyz$ for a standard LSST survey like day of observations, shown against the 1 s requirement.}\label{fig:settletime_xyz}
\end{figure}

Out of 487 slew events of the night starting on January 2nd 2026, 26 had at least one failure (Figure \ref{fig:settletime_xyz}), defined as one of the position or rotation columns from the IMS being above the repeatability requirement as measured by the IMS at any point between the 1 second mark after the slew stop and 15 seconds after it.

\begin{figure}[htbp]
    \centering
    \includegraphics[width=0.6\linewidth]{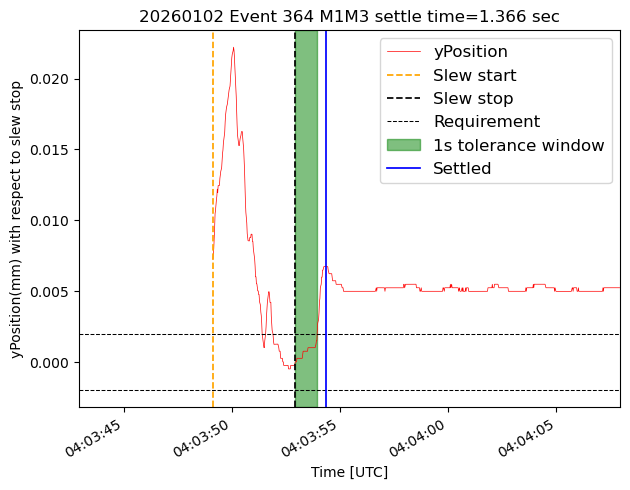}\caption{Example of a settling time failure in $y$ position. The $y$ axis is the one exposed to larger stresses from gravitational loads.}\label{fig:settletime_fail}
\end{figure}
\begin{figure}[htbp]
    \centering    \includegraphics[width=0.8\linewidth]{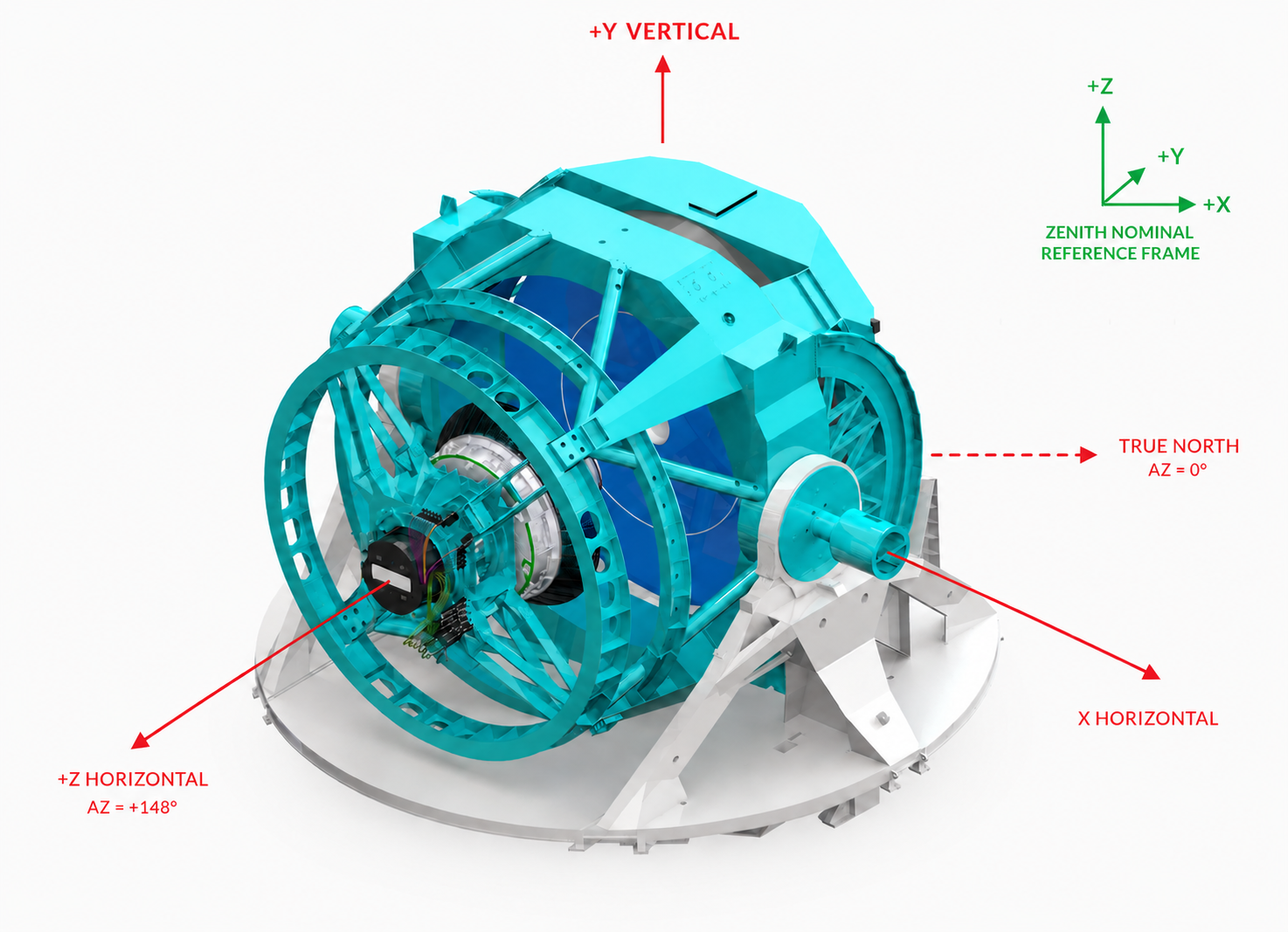}\caption{Reference axes for the telescope system.}\label{fig:telescope_axes}
\end{figure}

Failures to settle in the allocated time of one second after the slew stop are systematically on the order of $5\mu m$ shifts, just above the requirement in the $y$ direction (Figure \ref{fig:settletime_fail}). This is the direction in which one would expect any sort of deviation, as it lies in the gravitational $yz$ plane of the telescope, large loads have to be compensated with off-axis movements by the force actuators (see Figure \ref{fig:telescope_axes} for a reference). Nonetheless, in general, the amount of failures has been reduced drastically since the tests performed with the M1M3 surrogate (approximately from $25\%$ to $5\%$ and currently at a higher configuration speed for the TMA of $20\%$). It should be noted that the actual requirement is the 5 second total slew and settle time. The 1 second M1M3 setting time is a derived requirement, and the 5 second actual requirement can be met with a shorter slewing time and a longer settling time. 

\subsection{Braking distance}\label{sec:braking_distance}

To minimize heat production and component wear, deceleration is normally accomplished with the drive system. A separate traditional mechanical brake system is used for emergency braking. The mechanical brake test deliberately interrupts the TMA slew in order to evaluate the safety performance of the braking system. Specifically, it assesses the response time of the TMA to the brake command and quantifies the stopping distance, defined as the distance traveled from the instant the brake is activated until the TMA reaches a complete standstill. We used two ways to give a brake to the telescope, one is using the E-stop, this was done when the M1M3 surrogate mirror was installed on the telescope. The other one is to use the GIS door interlock - this was done with the real M1M3 mirror during the ComCam on-sky campaign, and the LSSTCam pre-LSST early on-sky operations. This test is done with emergency brake mechanical brake, not the normal slew and settle brake. 

Table \ref{tab:az_el_braking_distances} shows the stopping distances with different speed settings and with different systems installed. All cases were considered to be safe to run TMA in that speed settings in respect to braking distance. As expected, the braking distance is polynomially proportional to the TMA speed, as shown in the Figure \ref{fig:stopping_distance}. 

\begin{table}[htbp]
\centering
\begin{tabular}{l c c c c l}
\toprule
Case & Axis & Speed (\% of Max)& Speed (deg/s) & Stopping distance (deg) & Notes \\
\midrule

1-1  & Az & 70\% & 7.0 & 12.87 & M1M3 surrogate, 2023-11-23\\
1-2  & El & 70\% & 3.5 & 4.46  & M1M3 surrogate, 2023-11-23\\
\midrule
2-1  & Az & 1\% & 0.1 & 0.091 & ComCam on‑sky, 2024-11-06 \\
2-2  & El & 1\% & 0.1 & 0.12  & ComCam on‑sky, 2024-11-06 \\
\midrule
2-3  & Az & 2\% & 0.2 & 0.049 & ComCam on‑sky, 2024-11-08 \\
2-4  & El & 2\% & 0.2 & 0.073 & ComCam on‑sky, 2024-11-08 \\
\midrule
2-5  & Az & 5\% & 0.5 & 0.11  & ComCam on‑sky, 2024-11-09 \\
2-6  & El & 5\% & 0.5 & 0.22  & ComCam on‑sky, 2024-11-09 \\
\midrule
2-7  & Az & 10\% & 1.0 & 0.51  & ComCam on‑sky, 2024-11-21 \\
2-8  & El & 10\% & 1.0 & 0.60  & ComCam on‑sky, 2024-11-21 \\
\midrule
2-9  & Az & 20\% & 2.0 & 1.6   & ComCam on‑sky, 2024-11-29 \\
2-10 & El & 20\% & 2.0 & 1.91  & ComCam on‑sky, 2024-11-29 \\
\midrule
2-11 & Az & 40\% & 4.0 & 4.88  & ComCam on‑sky, 2024-12-08 \\
2-12 & El & 40\% & 4.0 & 5.95  & ComCam on‑sky, 2024-12-08 \\
\midrule
2-13 & Az & 20\% & 2.0 & 1.2  & LSSTCam pre-LSST, 2025-11-16 \\
2-14 & El & 20\% & 2.0 & 1.6  & LSSTCam pre-LSST, 2025-11-16 \\
\bottomrule
\end{tabular}
\caption{Stopping distance vs speed in Az and El braking tests}
\label{tab:az_el_braking_distances}
\end{table}

\begin{figure}[htbp]
    \centering
    \begin{subfigure}[b]{0.8\textwidth}
        \centering
        \includegraphics[width=0.8\linewidth]{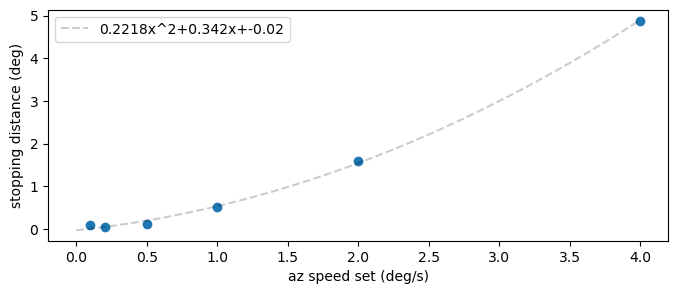}
        \caption{Stopping distance vs TMA azimuth speed}
    \end{subfigure}
    \begin{subfigure}[b]{0.8\textwidth}
        \centering
        \includegraphics[width=0.8\linewidth]{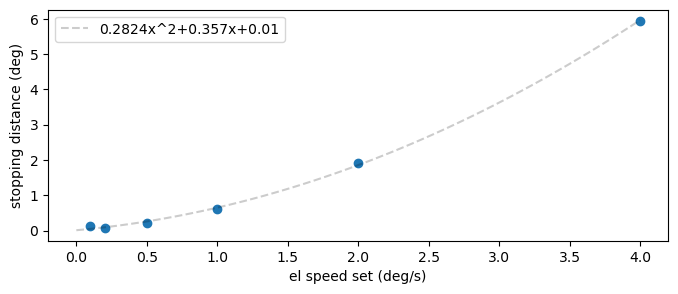}
        \caption{Stopping distance vs TMA elevation speed}
    \end{subfigure}
    \caption{Braking distance is polynomially proportional to the TMA speed.}
\end{figure}\label{fig:stopping_distance}

\newpage
\section{QUASI-STATIC TESTS}\label{sec:quasi-static}

Before raising M1M3, the force actuators (FA) and hardpoints (HP) are tested daily to verify that they are functioning properly for safe operation. More details on these two tests, the bump test and hardpoint breakaway test, can be found in in Bruno Q. et al. (2024)\cite{10.1117/12.3019268}. In this paper, we focus on the updates since October 2024.

\subsection{Bump test}
To test the functionality of the pneumatic actuators on the M1M3 mirror cell, small (222 N) perturbations are applied to each in turn, for a few seconds. This is called the bump test. 

\begin{figure}[htbp]
    \centering
    \includegraphics[width=0.8\linewidth]{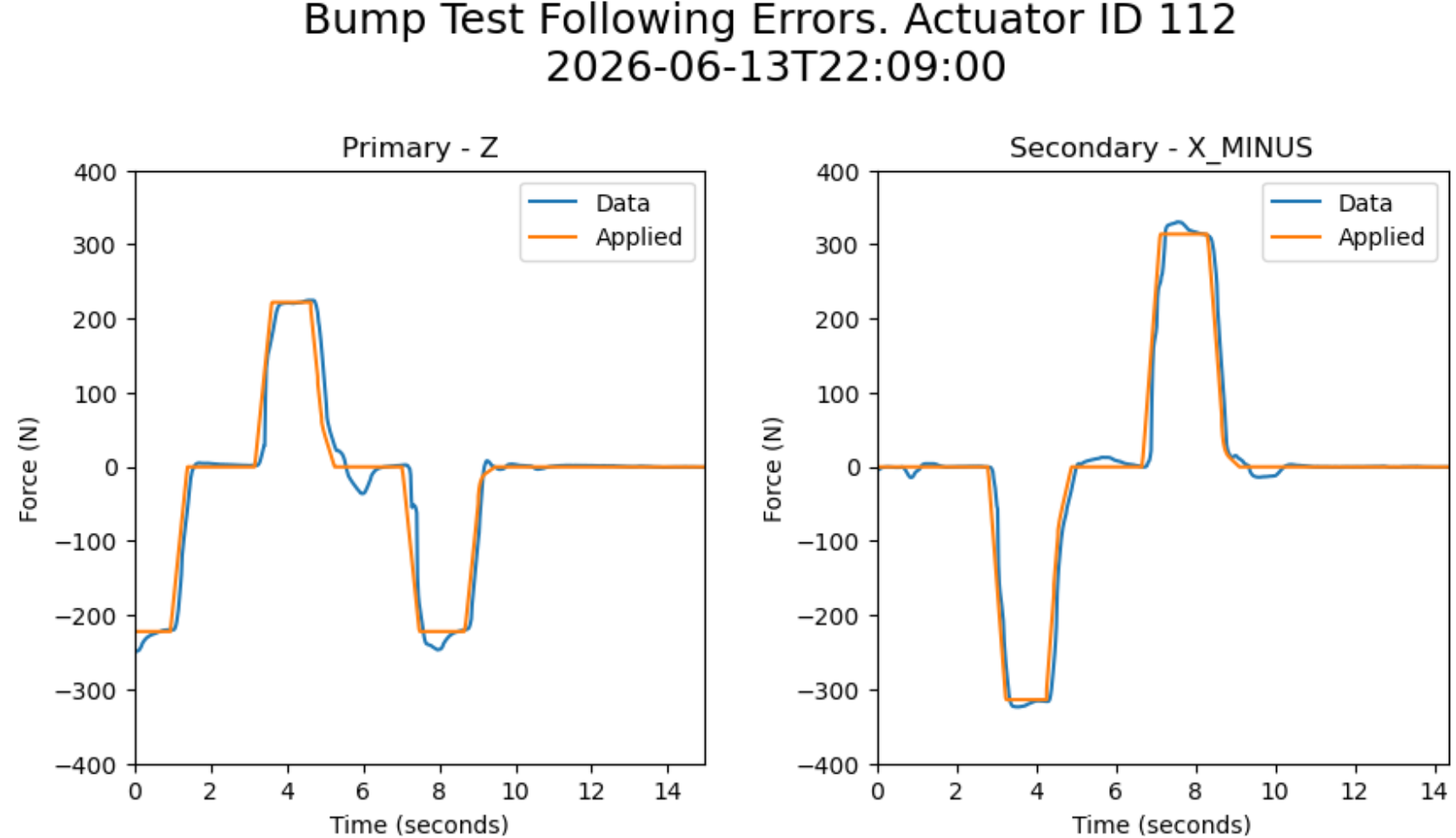}
    \caption{Example of the bump test on actuator 112, both on Z and X axes.}
    \label{fig:bump_test}
\end{figure}

Figure~\ref{fig:bump_test} shows an example of the applied (target) and actual measured forces on a given force actuator with two axes. 

\begin{figure}[htbp]
    \centering
    \includegraphics[width=\linewidth]{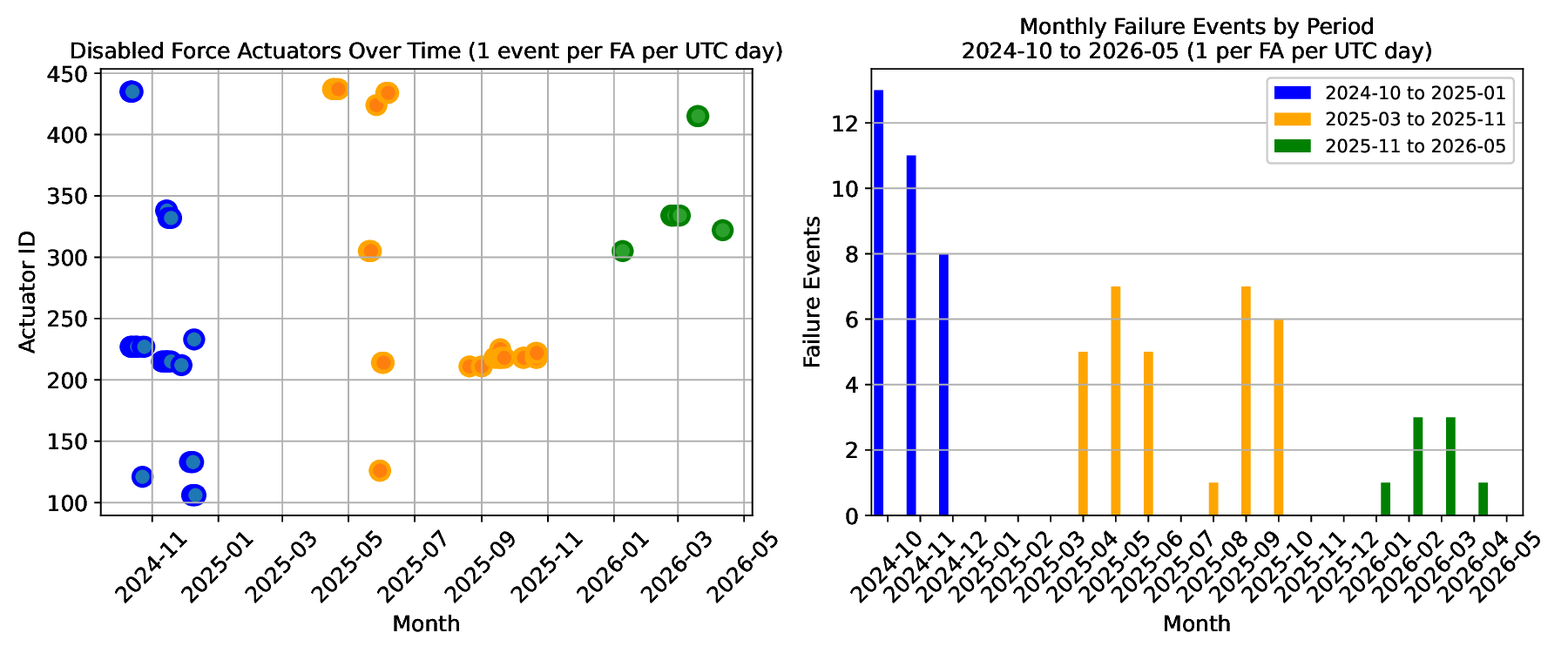}
    \caption{Histogram of failing FAs during three periods. Period 1: 2024 October to 2025 March (Blue, ComCam On-sky). Period 2: 2025 March to 2025 November 25 (Yellow, after LSSTCam was installed until the valves were replaced). Period 3: 2025 November 26 to Now (Green, since the valve replacements and dry-out).}\label{fig:fa_fail}
\end{figure}

There were a few force actuators that failed frequently and consistently. We found signs in them of internal corrosion caused by humidity in the valves. Since our air lines have dryers, we expected that ambient moisture may have entered the actuators while they were removed from the cell (during the period when we swapped the surrogate for the glass mirror). When we found evidence of corrosion, we replaced a substantial portion of the valves and put a cap over the pipes that would be in an open humid environment. These histograms in Figure~\ref{fig:fa_fail} compare the test performance before and after the corrosive valves replaced and kept dry. 

\begin{figure}[htbp]
    \centering
    \includegraphics[width=0.5\linewidth]{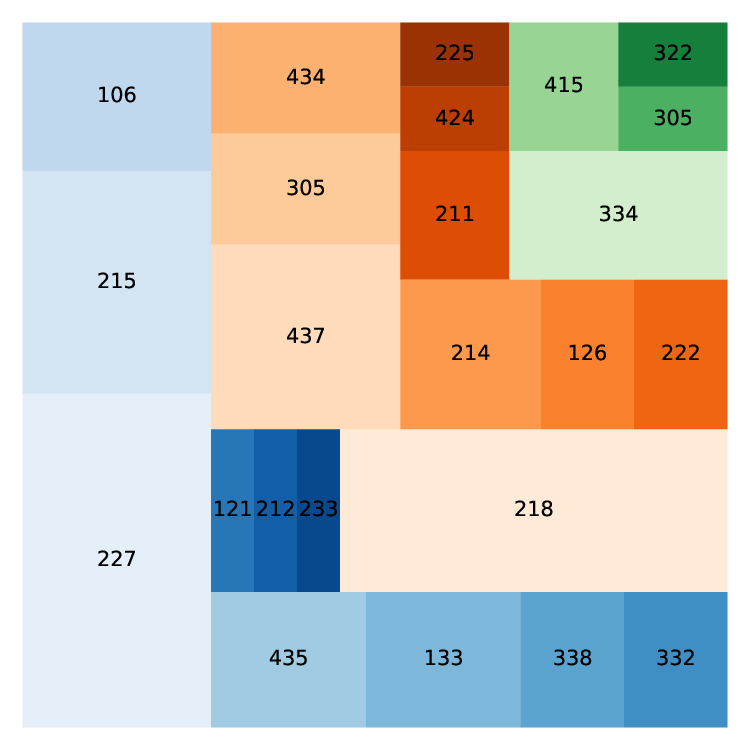}
    \caption{Treemap of failing FAs during three periods. The sequential colours indicate Period 1: 2024 October to 2025 March (Blue, ComCam On-sky). Period 2: 2025 March to 2025 November 25 (Yellow, after LSSTCam was installed until the valves were replaced). Period 3: 2025 November 26 to Now (Green, since the valve replacements and dry-out). The inset numbers are the FA IDs and the sizes of each rectangle scale according to the sum of unique daily failures in the target time period.}\label{fig:fa_tree}
\end{figure}

While we were testing the M1M3 system, after changing from the surrogate mirror to the glass mirror, software changes were also introduced to reduce false error messages on FAs that are actually functioning adequately. After installing the LSSTCam (yellow bars in Figure~\ref{fig:fa_fail}), we noticed that FAs were failing at the same location repeatably even after replacing the failing FAs. Then after finding the evidence of water residue in the FAs and pipes near these frequently failing FAs. On November 26, valves were replaced and kept dry, and we could see significant improvement on the FAs' behavior.

From 2024 October to 2025 January, there were 32 total events of FAs failing, meaning 8 events per month. From 2025 March to 2025 November 25th there were 31 total events, 3.4 events per month. After 2025 November 26th to now (2026 May 12th), there have been only 8 events, and it was mostly on two actuators that were replaced along with the connector and valve replacement near them, on March. Figure ~\ref{fig:fa_tree} demonstrates significant improvement in the FA Bump tests across the three periods -- the total number of failed FAs and the associated the total number of unique daily failures improve after the valve replacement. It is also worth noting that each of the three periods present unique failed FAs indicating overall progress in the regular verification of the M1M3 FAs. 


A major software enhancement was the parallelization of the bump tests. This modification enables testing of up to four force actuators (FAs) simultaneously, thereby substantially reducing the total time required to complete bump tests for all actuators. In combination with additional optimizations, the duration of the full bump test sequence for all FAs was reduced from more than 90 minutes to approximately 20 minutes. To the best of our knowledge, this improvement represents the first implementation of such large-scale parallelized bump testing on comparable deformable or active mirrors. As a result, the summit operations team gains additional time to prepare the observatory for nighttime observations following the handover of observatory subsystems from daytime maintenance activities.

\subsection{Hardpoint Breakaway Test}
\begin{figure}[htbp]
    \centering
    \includegraphics[width=0.8\linewidth]{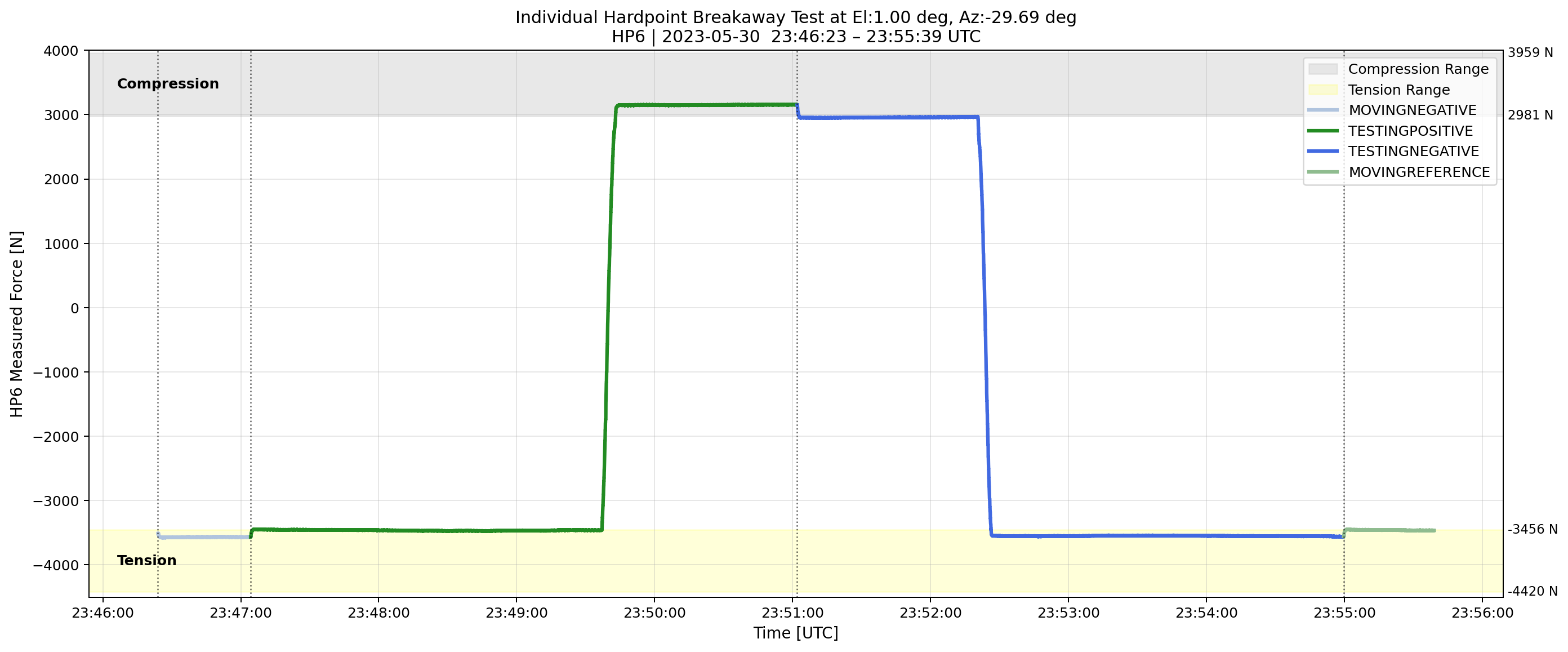}
    \caption{Transition of the measured forces on each hardpoint when the TMA is at el = 1deg}\label{fig:HP_breakaway_sample}

\end{figure}

Another test the observatory runs daily to test the M1M3 system before observation is the hardpoint breakaway test. The active support system of the M1M3 includes six axial hardpoint actuators in a hexapod configuration. Forces on these hardpoints are minimized during slewing by the dynamic offloading by the FAs.  The hardpoints have force limiting pneumatic breakaway mechanisms, which loose there stiffness when their breakaway limits are reached, greatly limiting the ability to transmit large forces. The breakaway limit for each hardpoint is in the range of -4420 N to -3456 N for retraction and 2981 N to 3959 N for extension. For this test, we move the hardpoints in negative (increasing tension) direction until the breakaway mechanism activates, then move the hardpoint in positive (increasing compression) direction until the breakaway mechanism activates as shown in the Figure. This test demonstrates that these breakaway mechanisms are functioning safely and properly\ref{fig:HP_breakaway_sample}.

\begin{figure}[htbp]
    \centering
    \begin{subfigure}[b]{\textwidth}
        \centering
        \includegraphics[width=\textwidth]{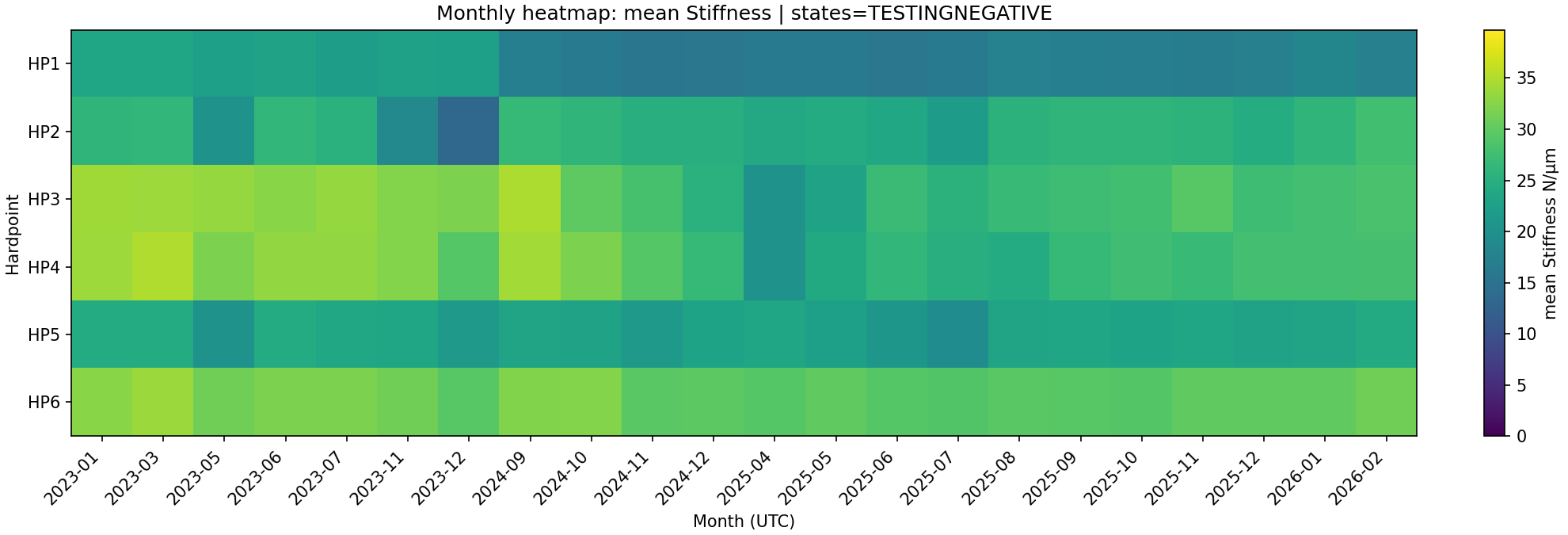}
        \caption{Heatmap of monthly stiffness mean (negative/compression state).}
    \end{subfigure}
    
    \begin{subfigure}[b]{\textwidth}
        \centering
        \includegraphics[width=\textwidth]{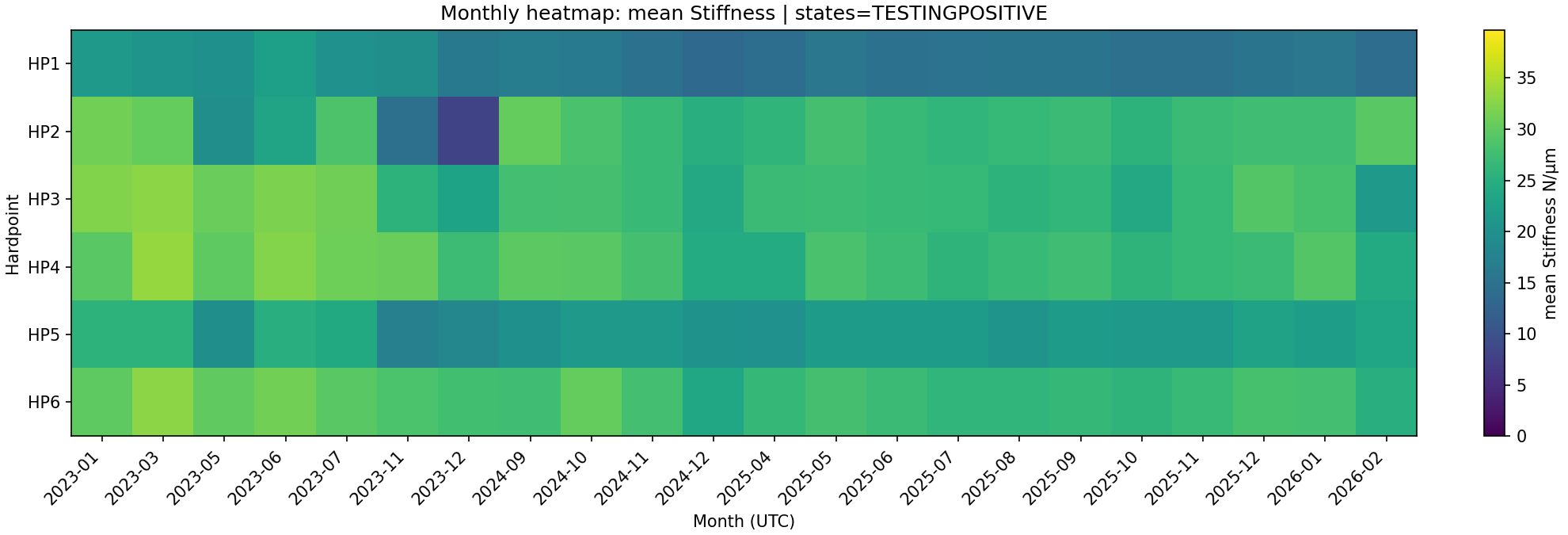}
        \caption{Heatmap of monthly stiffness mean (positive/tension state).}
    \end{subfigure}
    \caption{Both plots show high stiffness values during 2023 and 2024}
    \label{fig:heatmap}
\end{figure}

The breakaway test is also used to monitor the system stiffness of the hardpoints, which is critical to operations. The historical analysis of the M1M3 breakaway hardpoint tests shows that the period from 2023 to late 2024 was the most unstable, whereas the period from 2025 to the present shows a more linear behavior as the M1M3 glass mirror was installed. The heatmap in Figure \ref{fig:heatmap} indicates that HP2 exhibited lower stiffness at the end of 2023. Data collected prior to 2025 is considered non-representative. From 2025 onward, the data is reliable and could be used for behavioral modeling. 

\begin{figure}[htbp]
    \centering
    \begin{subfigure}[b]{0.7\textwidth}
        \centering
        \includegraphics[width=\textwidth]{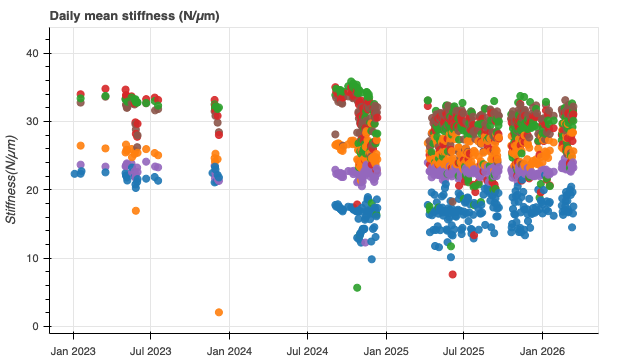} 
        \caption{Testing negative}
    \end{subfigure}
    \hfill 
    \begin{subfigure}[b]{0.7\textwidth}
        \centering
        \includegraphics[width=\textwidth]{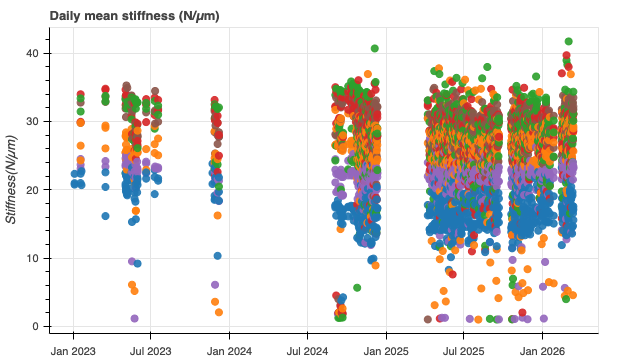} 
        \caption{Testing positive}
    \end{subfigure}
    \vspace{10pt} 
    \caption{Dashboard of hardpoint stiffness trends since January 2023 to April 2026. HP1: blue, HP2: orange, HP3: green, HP4: red, HP5: purple, HP6: brown.}
    \label{fig:history_stiffness}
\end{figure}

Figure~\ref{fig:history_stiffness} shows the daily mean stiffness evolves over time, modeled by HP and state. In general, we observe that stiffness values tend to become more linear over time.

For elevation angles from 0 to 70 degrees, there were insufficient data points for each HP to calculate the median or standard deviation via normal fit, given our minimum requirement of n$\ge$10. For elevation angles above 70°, we observe that HP1 reaches a minimum mean stiffness of 15.02 N/µm in the 80°–85° range (=2.38), while HP6 reaches a maximum of 31.03 N/µm in the 85°–90° range (=1.78). These are the system level stiffnesses that include the combined effects of the hardpoint, mirror and mirror cell, and not the actual stiffness of the hardpoint itself. Overall, a general trend is observed that for each hardpoint, an increase in elevation angle correlates with a decrease in stiffness values. The range over which a HP can be evaluated can be extended by adjusting its length. We intend to implement these modifications as time permits.

HP2 and HP5 exhibit higher instability at elevation angles below 60°. While this behavior was clearly visible in pre-2025 datasets, there is currently insufficient data in the recent logs to confirm if this trend persists.

\begin{figure}[htbp]
    \centering
    \includegraphics[width=0.7\linewidth]{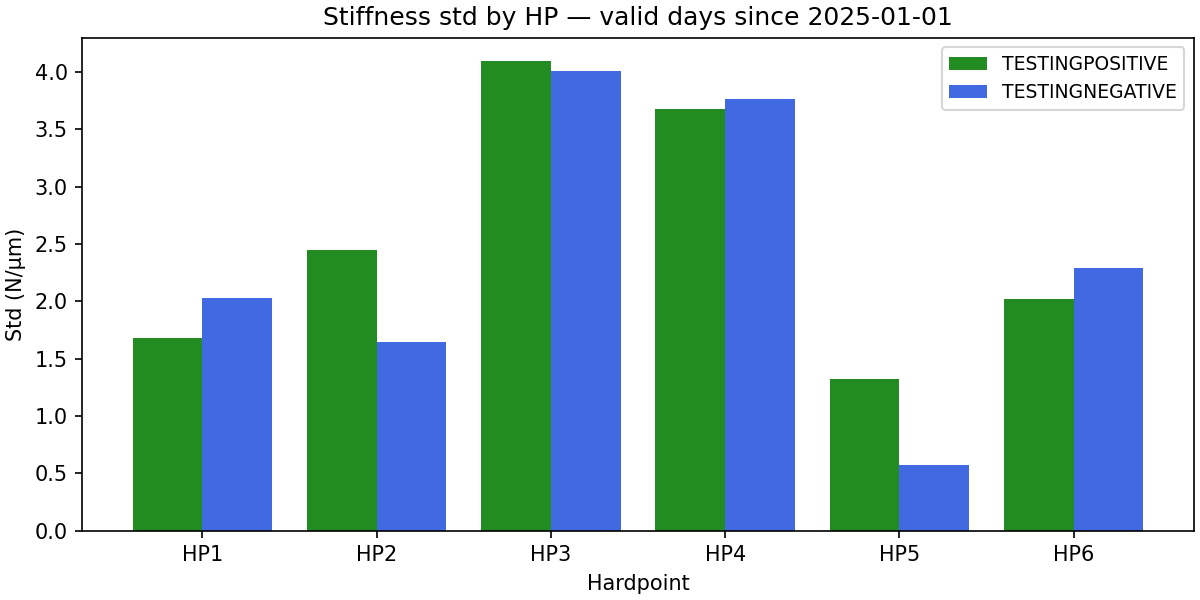}
    \caption{Standard deviation for each hp since 2025.}\label{fig:HP3HP4}
\end{figure}

HP3 and HP4 show significantly higher dispersion and fluctuations compared to the other hardpoints as shown in Figure~\ref{fig:HP3HP4} and Table~\ref{tab:hp-sigma}. 

\begin{table}[ht]
\centering
\begin{tabular}{c c c}
\toprule
HP & $\sigma$ Testing Negative & $\sigma$ Testing Positive \\
\midrule
1 & 2.032 & 1.677 \\
2 & 1.646 & 2.445 \\
3 & 4.012 & 4.093 \\
4 & 3.764 & 3.673 \\
5 & 0.573 & 1.321 \\
6 & 2.288 & 2.024 \\
\bottomrule
\end{tabular}
\caption{Standard deviation for each HP for state.}
\label{tab:hp-sigma}
\end{table}

\section{RESPONSE TO THE EARTHQUAKE}\label{sec:earthquake}
The Vera C. Rubin Observatory is located in the Coquimbo Region of Chile on the mountaintop of Cerro Pachón in the Andes Mountains with latitude -30.24465$^o$ and longitude -70.74934$^o$. The site is located in a high-seismicity region. Therefore we have to monitor the earthquakes and need the Global Interlock System (GIS) to response to moderate to major earthquakes. There were 113 earthquakes detected since September 2025. Table \ref{tab:earthquake} shows eleven earthquake events which had more than magnitude 5.0, and two of them were over 6.0 magnitude. The Rubin Observatory's dome lower enclosure is slightly separated from the telescope's pier so they can shake independently without crushing each other. 

The mirrors of the Rubin Observatory are engineered to actively respond to external mechanical perturbations in order to adapt to seismic and other dynamic loads. This adaptive behavior is intended to protect the optics from damage while maintaining continuous nighttime operations under small- to moderate-magnitude shocks, without interrupting observations or compromising mirror integrity. In the event of a major earthquake, however, the M1M3 mirror system is designed to be transferred to its static support configuration within 2 seconds of the onset of seismic activity. In this configuration, more than 400 wire rope isolators, which constitute the static support system, fulfill one of their primary functions: dissipating and redistributing the seismic energy—together with the loads transmitted by the mirror positioning system—so that the monolithic mirror assembly preserves its structural integrity and remains as a single, undamaged solid element.

\begin{table}[htbp]
\centering
\begin{tabular}{l c c r r c c r}
\toprule
Timestamp (UTC) & Mag. & Latitude & Longitude & Distance & State & Reaction & Peak force \\
\midrule

2025-10-02 10:43:55 & 5.4 & -30.6637$^o$ & -71.8514$^o$ & 115.46 km & static \\
2025-10-15 11:43:39 & 5.6 & -30.7205$^o$ & -71.5590$^o$ & 93.91 km & static \\
2025-11-03 15:51:54 & 5.5 &-27.4457$^o$ & -71.4784$^o$ & 319.22 km & static \\
2025-11-15 06:13:57 & 5.0 &-28.7985$^o$ & -71.5292$^o$ & 177.63 km & active & damped & $<$800 N \\
2025-12-10 06:38:27 & 5.2 &-28.8659$^o$ & -71.3475$^o$ & 163.86 km & active & damped & $<$2000 N \\
2026-01-08 14:14:55 & 5.2 &-31.2688$^o$ &-67.5932$^o$ & 322.35 km & active & damped & $<$200 N\\
2026-02-12 13:34:31 & 6.2 &-30.8012$^o$ &-71.4451$^o$ & 90.95 km & static \\
2026-03-07 17:13:26 & 5.1 &-31.6647$^o$ & -67.0247$^o$ & 388.66 km & active & damped & $<$200 N \\
2026-03-09 06:10:17 & 5.1 &-31.4461$^o$ &-69.5425$^o$ & 176.41 km & active & damped & $<$2000 N \\
2026-03-13 13:39:18 & 6.3 &-28.6840$^o$ &-71.6269$^o$ & 193.22 km & active & oscillates & $<$4000 N \\
2026-04-16 03:24:06 & 4.0 &-30.2486$^o$ &-71.2004$^o$ & 43.33 km & active & faulted & $<$2400 N \\
2026-04-24 13:49:41 & 5.0 &-27.9200$^o$ &-70.9962$^o$ & 259.60 km & static \\
2026-06-02 06:13:42 & 4.5 &-29.3593$^o$ &-70.9171$^o$ & 99.77 km & active & faulted & $<$1500 N \\
\midrule
\end{tabular}
\caption{Earthquake with magnitude larger than 4.5 since 2025 September. Maximum ground-motion amplitudes at earthquake epicenters as reported by the U.S. Geological Survey (USGS).}\label{tab:earthquake}
\end{table}

The maximum allowable force transmitted by the HP to ensure the structural integrity of the mirror during a magnitude 6 earthquake is ideally remain below 1500 N as a software limit. To investigate this condition, it is necessary to analyze the forces exerted by the HP along the corresponding axes and compare them with the associated accelerations.

\begin{figure}[bp]
    \centering
    \includegraphics[width=0.8\linewidth]{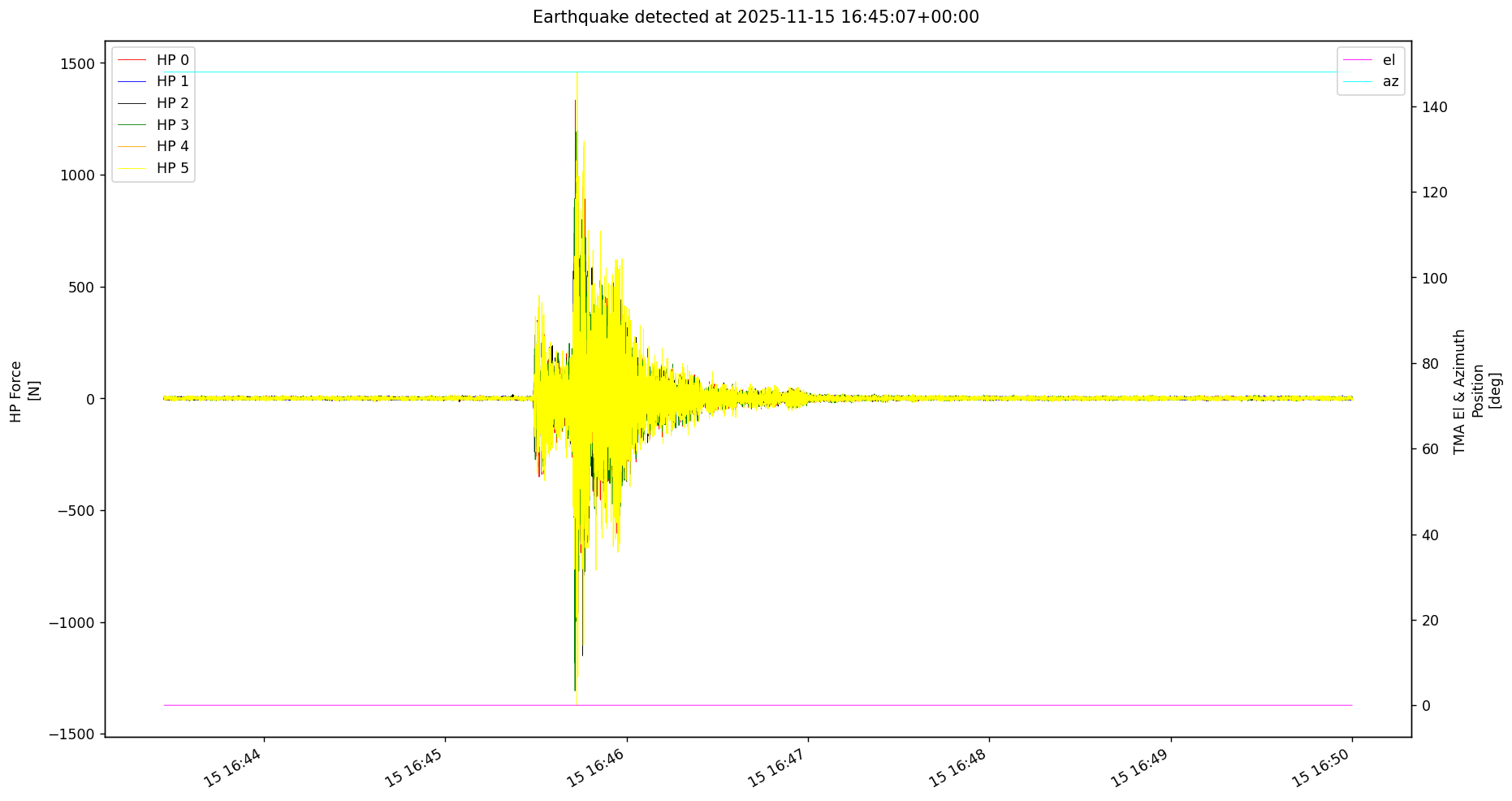}
    \caption{Example of M1M3 HP forces when earthquake happened in 2025 November 15.}\label{fig:20251115_earthquake}

\end{figure}

On February 12th and March 13th, earthquake with magnitude over 6.0 occurred. We were able to observe minor effects of the earthquake on the mirror on February 12th, as the mirror was resting on static support. Detailed analysis of the TMA reaction on February 12th is provided in Freddy Muñoz Arancibia et.al (2026)\cite{balancing}.

\subsection{March 2026 Earthquakes}

Both magnitude‑five seismic events in March 2026 occurred while the mirror was in the raised position (Supported by the FAs and HBs, rather than the static supports), and in each case the mirror support system successfully damped, thereby preventing any significant operational or structural concerns. The 9th March event took place while the TMA was slewing, a condition that may have contributed to partial absorption or redistribution of the earthquake-induced energy. 

\begin{figure}[bp]
    \centering
    \begin{subfigure}[b]{0.7\textwidth}
        \centering
        \includegraphics[width=0.8\linewidth]{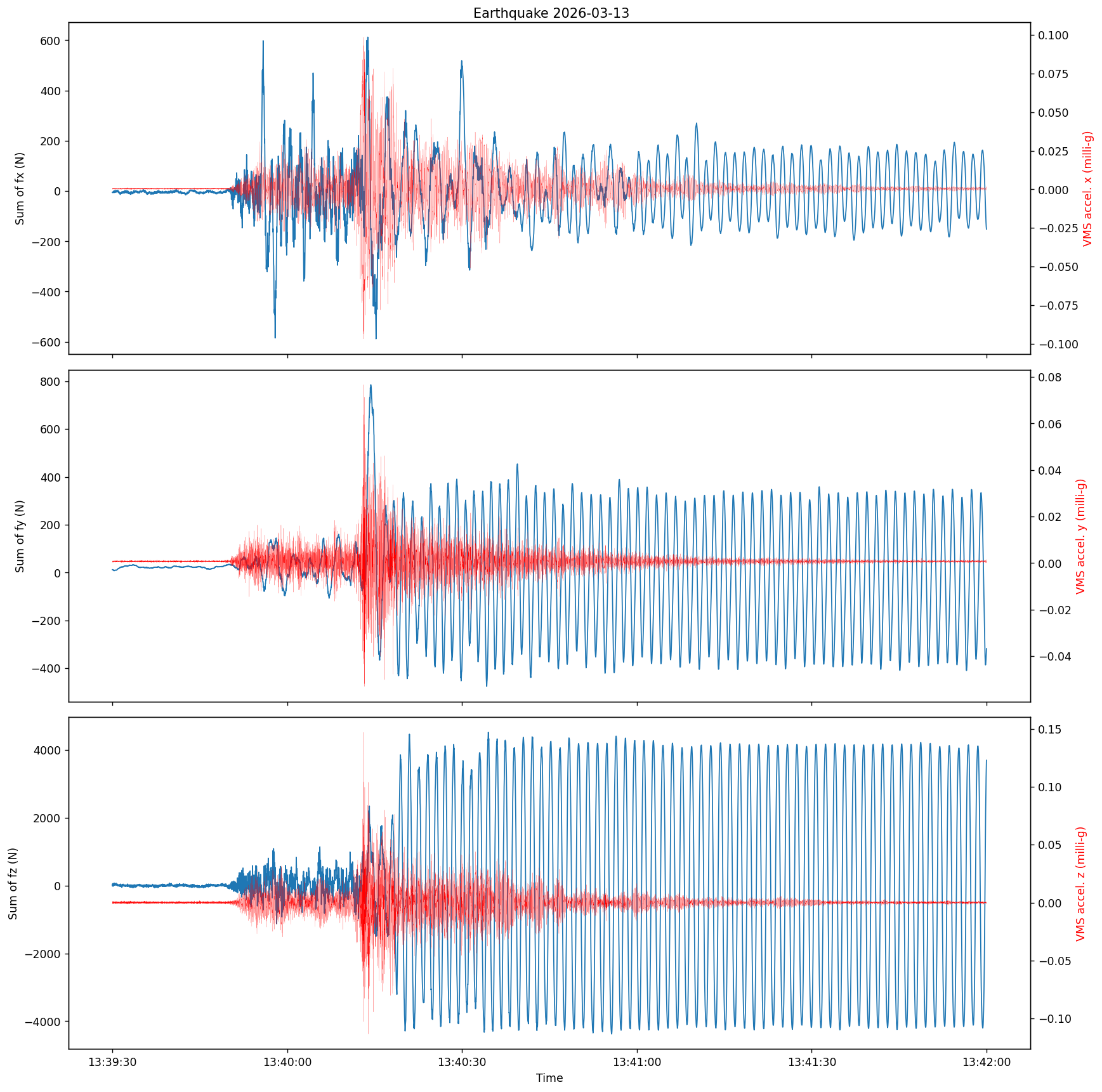}
        \caption{M1M3 applied forces compared with accelerometers on xyz axis, on March 13.}\label{fig:20260313_accel}
    \end{subfigure}

    \begin{subfigure}[b]{0.7\textwidth}
        \centering
        \includegraphics[width=0.8\linewidth]{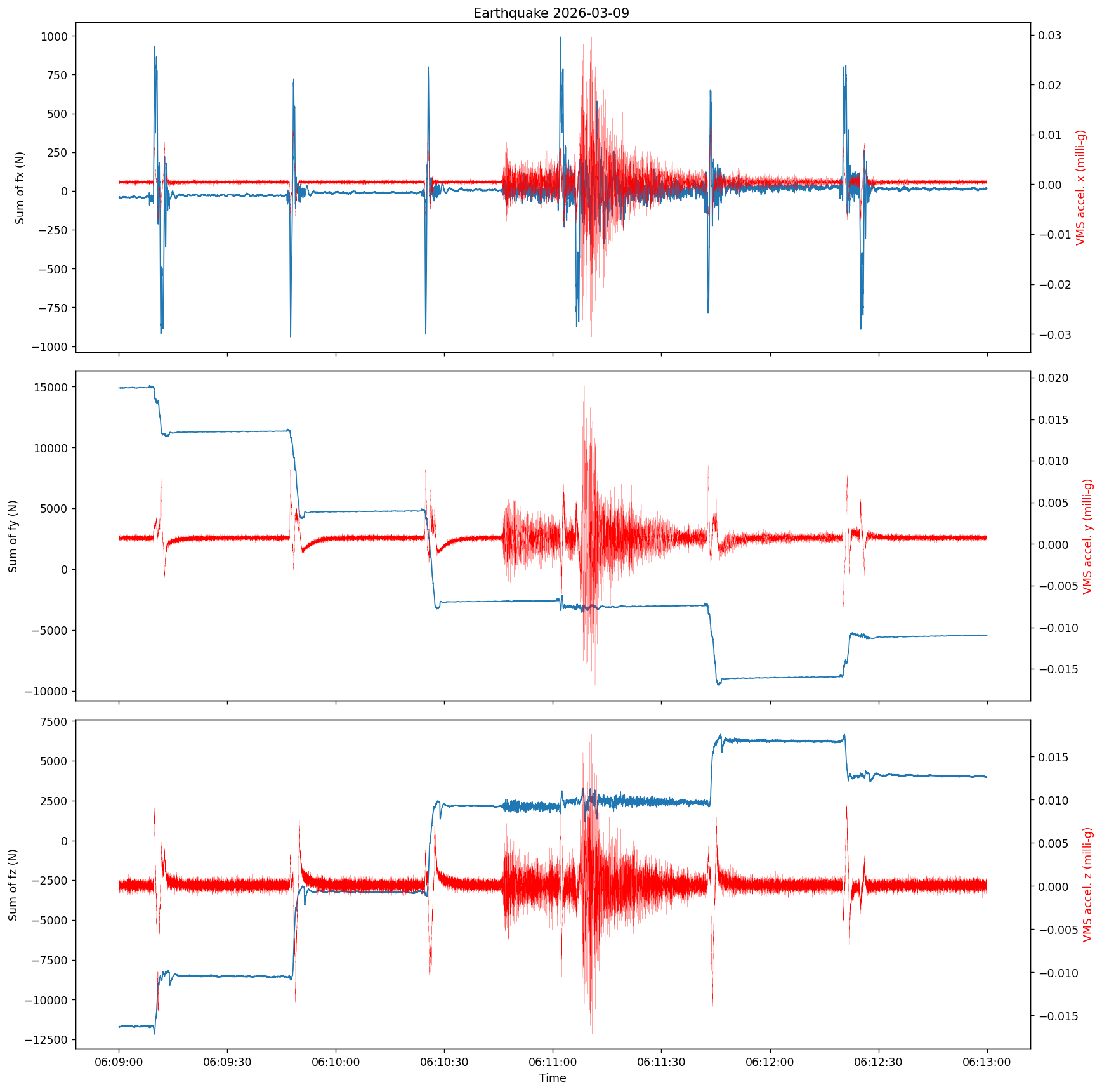}
        \caption{M1M3 applied forces compared with accelerometers on xyz axis, on March 09.}\label{fig:20260309_accel}
    \end{subfigure}
    \caption{Comparing two events of earthquakes that happened in March 13 (top) and March 09, 2026 to show the difference of the events that the mirrors weren't behaving properly after the earthquake to that the mirrors properly reacted to the earthquake and kept running the observations after that.}
\end{figure}

Comparing March 9th in Figure \ref{fig:20260309_accel} when the M1M3 was able to dampen the earthquake energy without entering sustained oscillation, with March 13 in Figure \ref{fig:20260313_accel}, when a much larger earthquake triggered large oscillations, suggests that something was wrong with the M1M3 control system.

\subsection{Updates after March 13th earthquake}
We adjusted the parameters to fault the mirror even for relatively mild earthquakes, provided they occurred close to the observatory. We need to collect more data to confirm this was a good move. We included in the table of two events when mirror was properly faulted during minor, but nearby events. As the profile shows forces which might lead to oscillations occurring on the March 13th 2026 event, we believe that was correct system behavior.

\begin{figure}[htbp]
    \centering
    \begin{subfigure}[b]{0.7\textwidth}
        \centering
        \includegraphics[width=0.8\linewidth]{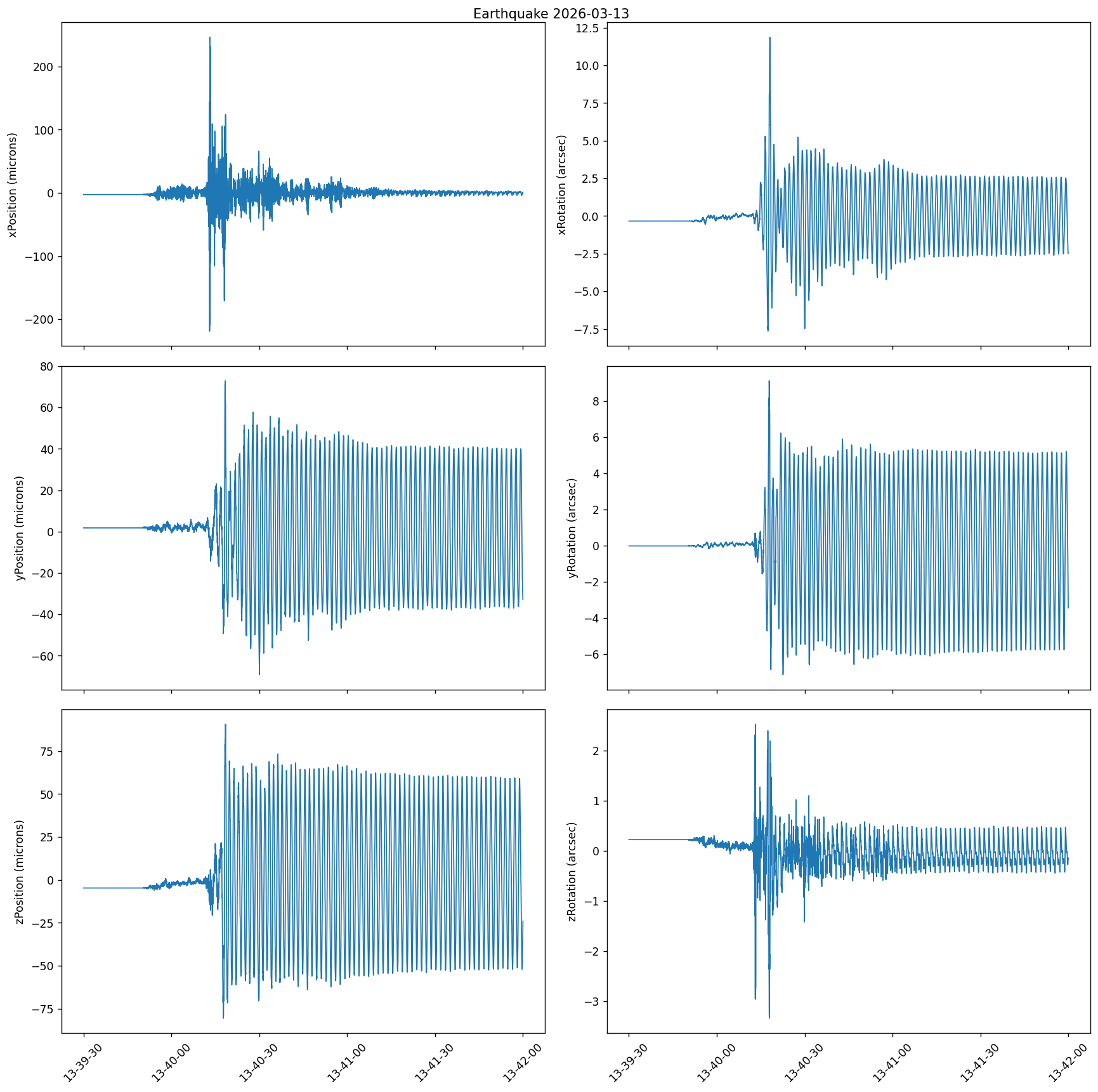}
        \caption{M1M3 position x, y, z, and rotation rx, ry, rz on March 13.}\label{fig:20260313_pos_rot}
    \end{subfigure}
    \begin{subfigure}[b]{0.7\textwidth}
        \centering
        \includegraphics[width=0.8\linewidth]{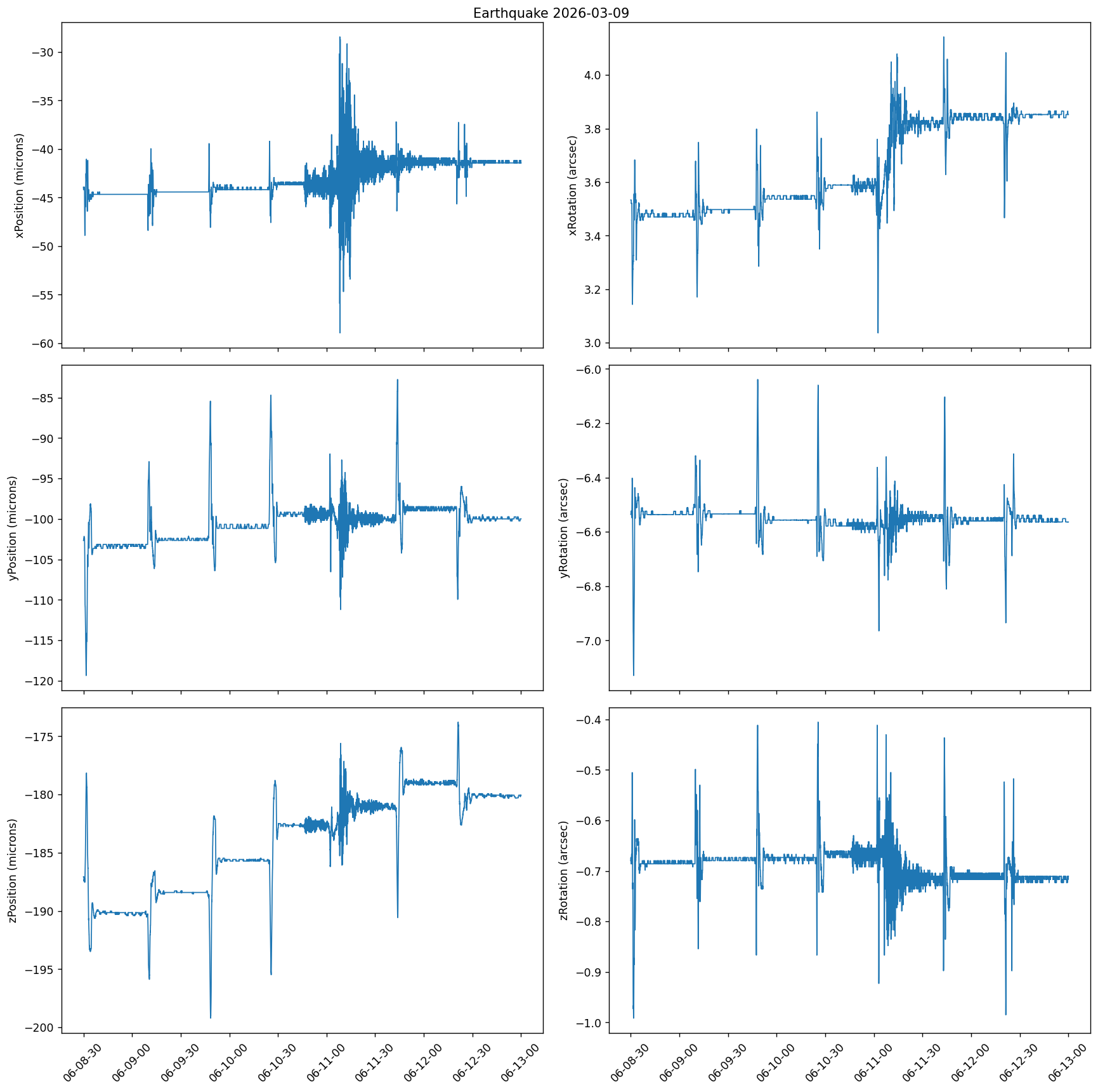}
        \caption{M1M3 position x, y, z, and rotation rx, ry, rz on March 09.}\label{fig:20260309_pos_rot}
    \end{subfigure}
    \caption{Comparing position xyz and rotation rxryrz after the earthquakes on March 13 when the reaction to the earthquake mechanism didn't work well (top) and March 09 (bottom) when mirrors reacted well and kept running the test as normal after that.}
\end{figure}

Also, comparing position x, y, z and rotation rx, ry, rz on March 9th in Figure \ref{fig:20260309_pos_rot}  to March 13th in Figure \ref{fig:20260313_pos_rot} we can see that the M1M3 kept oscillating, while big oscillation faded away and small vibrations happen time to time when slewing as normal, on March 9th event.

\section{CONCLUSION}
Since replacing the surrogate mirrors with the M1M3 in October 2024, we have conducted comprehensive testing—daily, routinely, and with each new system integration—to ensure the mirrors operate safely, properly, and stably.

As additional systems are installed on the TMA, we perform new balancing and update the lookup tables (LUTs) to account for the evolving mass distribution. We verified that slew and settle times meet specifications, confirmed the safety of operating the telescope with the mirrors at increased slewing speeds, and ensured repeatable positioning after slewing. Daily tests monitor the force actuators and hardpoints, with full historical tracking since the surrogate mirrors era. We have also tested and refined the mirrors’ response to earthquakes, enhancing safe and stable operations.



\acknowledgments 
We extend our sincere thanks to everyone involved in the summit for their tireless efforts in constructing these monumental systems collaboratively, troubleshooting and resolving failures, and operating the telescope and dome. The Vera C. Rubin Observatory is funded by the National Science Foundation (NSF) and the Department of Energy (DOE), whose generous support has been instrumental in advancing this groundbreaking project. We also greatly appreciate the invaluable remote support in software development, pipeline science, calibration, image analysis, and science data planning. Countless individuals worked in the middle of the night or staying up all night on the mountain to bring this project to fruition. 

Petr Kubánek used Gemini AI to calculate earthquakes distance and AI help available in Overleaf for text corrections.

\bibliography{report} 
\bibliographystyle{spiebib} 

\end{document}